\documentclass{jfm}
\usepackage{graphicx}
\usepackage{epstopdf, epsfig,color}
\usepackage{amssymb}
\usepackage{wrapfig,txfonts}
\usepackage{natbib}
\usepackage{url}

  \newcommand{\bfomega}{\mbox{\boldmath $\omega$}} 
 
   \newcommand{\be}{\begin{equation}}
      \newcommand{\ee}{\end{equation}}
         \newcommand{\tn}{\textnormal}
  \newcommand{\bea}{\begin{eqnarray}}
   \newcommand{\eea}{\end{eqnarray}}

\title[ Finite-time singularity]{Towards a finite-time singularity of the Navier-\\Stokes equations. Part 2. Vortex  reconnection\\ and singularity evasion }

\author[ H.K.Moffatt  and Y.Kimura ]{ H. K. Moffatt$^1$ and Yoshifumi Kimura$^2$ }

\affiliation{$^1$Department of Applied Mathematics and Theoretical Physics, \\
Wilberforce Road, Cambridge CB3 0WA, UK\\
[\affilskip]$^2$Graduate School of Mathematics, Nagoya University,\\
Furo-cho, Chikusa-ku, Nagoya 464-8602 Japan}

\pubyear{??} \volume{??} \pagerange{??--??}
\date{}
\begin{document}

\maketitle

\begin{abstract}
 In Part 1 of this work, we have derived a dynamical system describing the approach to a finite-time singularity of the Navier-Stokes equations. We now supplement this system with an equation describing the process of vortex reconnection at the apex of a pyramid, neglecting core deformation during the reconnection process. On this basis, we compute the maximum vorticity $\omega_{max}$ as a function of vortex Reynolds number $R_\Gamma$ in the range $2000\le  R_\Gamma \le 3400$, and deduce a compatible behaviour  
 $\omega_{max}\sim \omega_{0}\exp{\left[1 + 220 \left(\log\left[R_{\Gamma}/2000\right]\right)^{2}\right]}$ as  $R_\Gamma\rightarrow \infty$. This may be described as a physical (although not strictly mathematical) singularity, for all $R_\Gamma \gtrsim 4000$.
 \end{abstract}


\section{Introduction}\label{Sec_introduction}
 In Part 1 of this work (Moffatt \& Kimura 2019, hereafter MK19), we have derived a dynamical system governing the evolution of two initially circular vortices of radius $R$ and circulations $\pm\Gamma$. The vortices are symmetrically located on planes $x=\pm z \tan\alpha$, and are assumed to have Gaussian cores of initial radial scale $\delta_0$.  The Navier-Stokes equations are non-dimensionalised in terms of length-scale $R$ and time-scale $R^2/\Gamma$. The `tipping points'  are the points of closest approach of the vortices, and it turns out that the key variables are the separation $2s(\tau)$ of the tipping points, and the curvature $\kappa(\tau)$  and evolving radial scale  $\delta(\tau)$ at either tipping point, where $\tau=(\Gamma/R^{2})\,t$ is dimensionless time. It is assumed that 
\be\label{inequalities}
\delta(0)\ll s(0)\ll1,
\ee
and that the vortex Reynolds number $R_{\Gamma}=\Gamma/\nu$ is large (where $\nu$ is the usual kinematic viscosity of the fluid). The dynamical system derived in MK19 (eqns. 6.9a-c) is then
\be \label{system1}
\frac{\tn{d} s}{\tn{d}{ \tau}}=-\frac{\kappa\cos\alpha}{4\pi}\left[\log\left(\frac{s}{\delta}\right)+\beta_{1}\right] \,,\quad \frac{\tn{d} \kappa}{\tn{d}{\tau}}=\frac{\kappa\cos\alpha\sin\alpha}{4\pi {s}^2}\,,\quad \frac{\tn{d} \,\delta^2}{\tn{d}{\tau}}=\epsilon -\frac{\kappa\cos\alpha}{4\pi {s}} \,\delta^{2}\,,
\ee
where $\epsilon\equiv R_{\Gamma}^{-1}\ll 1$, and the parameter $\beta_{1}$ takes the value $\beta_{1}=0.4417$ for the assumed Gaussian vortex core structures. It was argued that the rate of strain in the neighbourhood of either tipping point is such as to turn the planes always towards angle $\alpha= \pi/4$, and that this constant value may therefore be adopted.  Equations (\ref{system1}) may then be integrated with initial conditions
\be
s(0)=s_0,\quad \kappa(0)=\kappa_0=1,\quad \delta(0)=\delta_0\,.
\ee

\begin{figure}
\begin{center}
\begin{minipage}{0.99\textwidth}
\hspace*{20pt}
(a)\includegraphics[width=0.45\textwidth, trim=0mm 0mm 0mm 0mm]{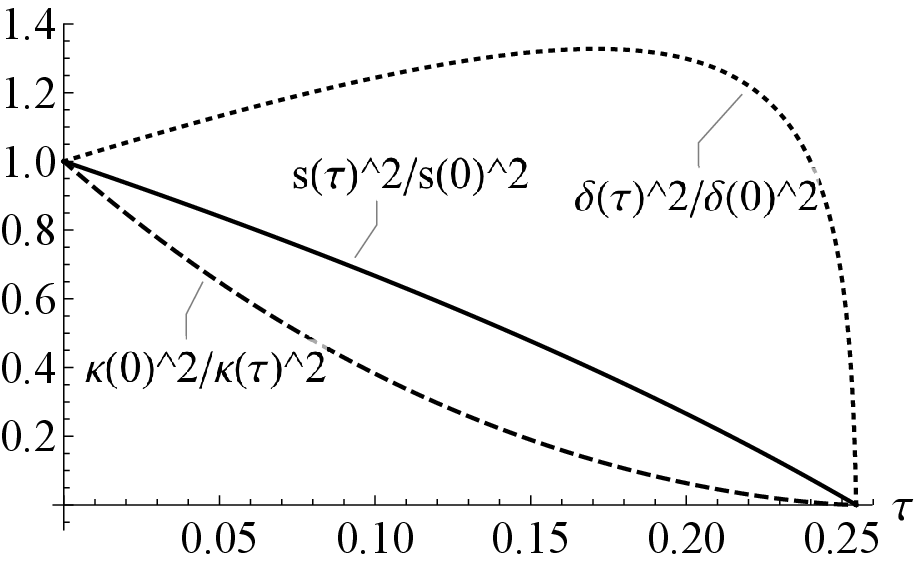}
\hspace*{10pt}
(b)\includegraphics[width=0.45\textwidth,  trim=0mm 0mm 0mm 0mm]{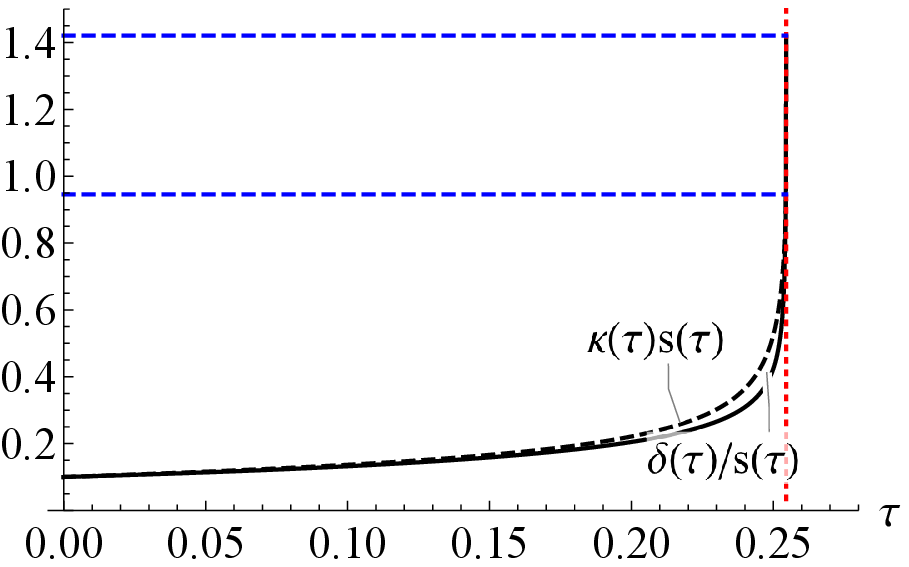}
\hspace*{20pt}
\vskip 2mm
\hspace*{20pt}
(c)\includegraphics[width=0.45\textwidth, trim=0mm 0mm 0mm 0mm]{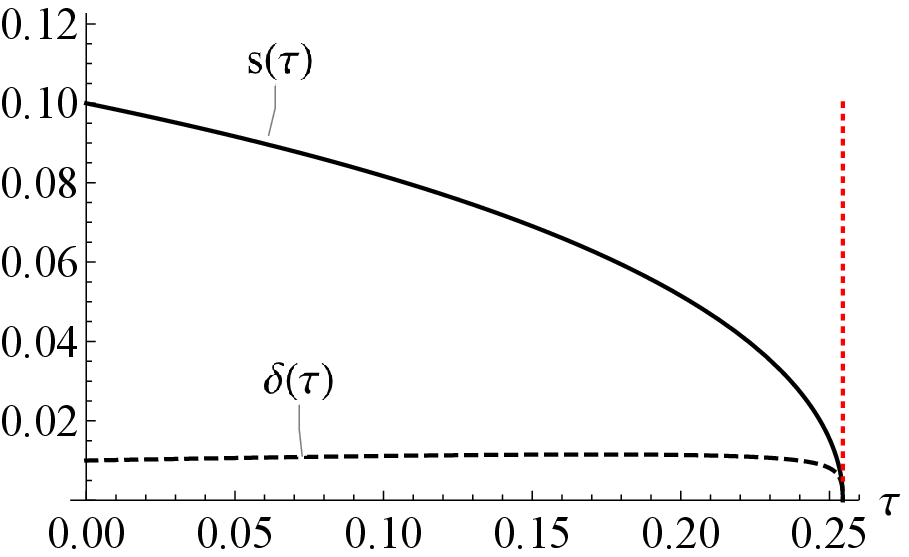}
\hspace*{10pt}
(d)\includegraphics[width=0.45\textwidth,  trim=0mm 0mm 0mm 0mm]{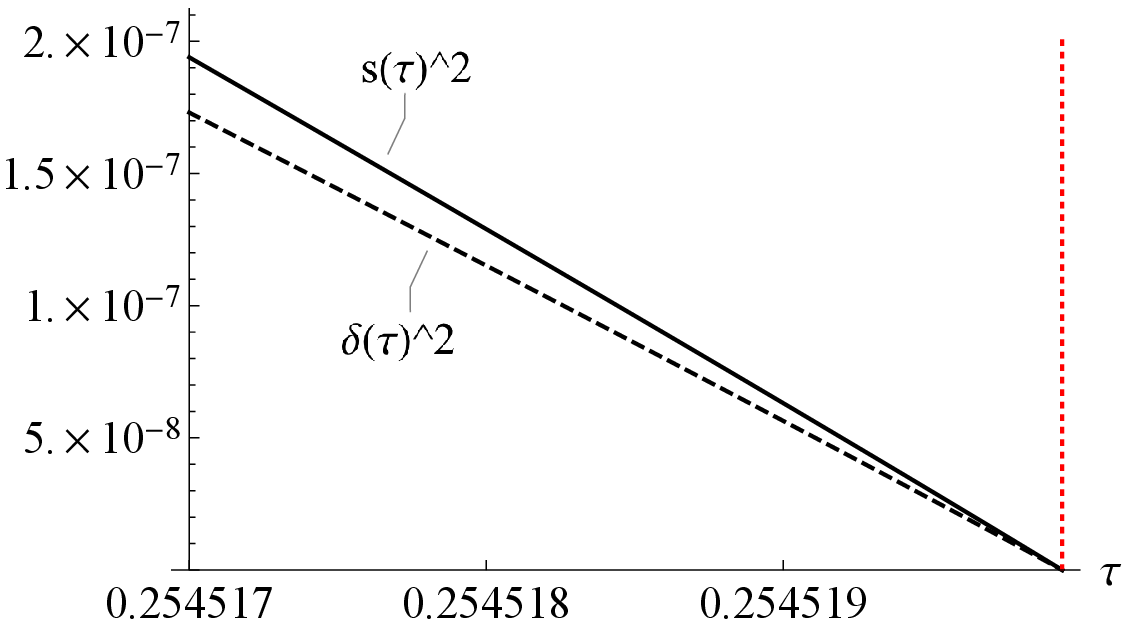}
\end{minipage}
\end{center}
\vskip 2mm
\caption{Evolution governed by  (\ref{system1}) with initial conditions (\ref{initial}) and with $\epsilon=1/3000$; \,(a) the three curves as labelled fall to zero at $\tau=\tau_c\approx 0.25452$;\, (b) the curves of $\kappa(\tau)s(\tau)$ (dashed) and $\delta(\tau)/s(\tau)$ (solid) rise to limiting values 1.4208 and 0.9456 respectively (blue, dashed) at  $\tau=\tau_c$ (red dotted);\, (c) $s(\tau)$ and $\delta(\tau)$, showing how both fall to zero at $\tau=\tau_c$; \, (d) very near $\tau_c$, where $s(\tau)^2$ and  $\delta(\tau)^2$ decrease linearly at constant ratio in the limit $\tau\rightarrow\tau_c$.}
\label{Fig_three_curves}
\end{figure}

In the present paper, we shall, by way of illustration, adopt the initial values
\be\label{initial}
s_0=0.1,\quad \kappa_0=1,\quad \delta_0=0.01.
\ee
This choice allows reasonable resolution of the behaviour near the critical time $\tau=\tau_c$ when consideration of vortex reconnection becomes essential.
(Other values compatible with (\ref{inequalities}) may be similarly treated, as done in MK19, where the case $s(0)=0.05,  \delta(0)=10^{-5},  \kappa(0)=1,  \epsilon=10^{-20}$ is described.  In all cases, it is found that, when $\epsilon \ll 1$, solutions of (1.2) become singular at a finite time $\tau_c$ that depends on $s(0)$, and that $\delta/s$ and $\kappa s$ approach asymptotic values as $\tau\rightarrow \tau_c$ near to the values $\surd{2}$ and $0.943367$ predicted by the asymptotic similarity solution obtained in \S 10 of MK19). \,Figure \ref{Fig_three_curves} shows the evolution of these variables from the initial conditions (\ref{initial}) with the choice $\epsilon=1/3000$, up to the critical time $\tau=\tau_c\approx  0.25452$. Figure  \ref{Fig_three_curves}b shows the corresponding variation of  $\kappa(\tau)s(\tau)$ and $\delta(\tau)/s(\tau)$, which rise in tandem to limiting values 1.4208 and 0.9456 respectively at  $\tau=\tau_c$. It is clear that when $\delta(\tau)/s(\tau)$ rises to about 0.2 and greater, the Biot-Savart law on which the first two equations of (\ref{system1}) are based requires modification to take account of inter-diffusion of the two vortices, and associated vortex reconnection.  It is the purpose of this follow-up paper to take due account of this effect, and to determine the resulting behaviour as the Reynolds number $R_\Gamma$ increases.
\section{The process of pyramid reconnection }\label{reconnection}
\begin{figure}
\begin{center}
\includegraphics[width=0.80\textwidth, trim=0mm 0mm 0mm 0mm]{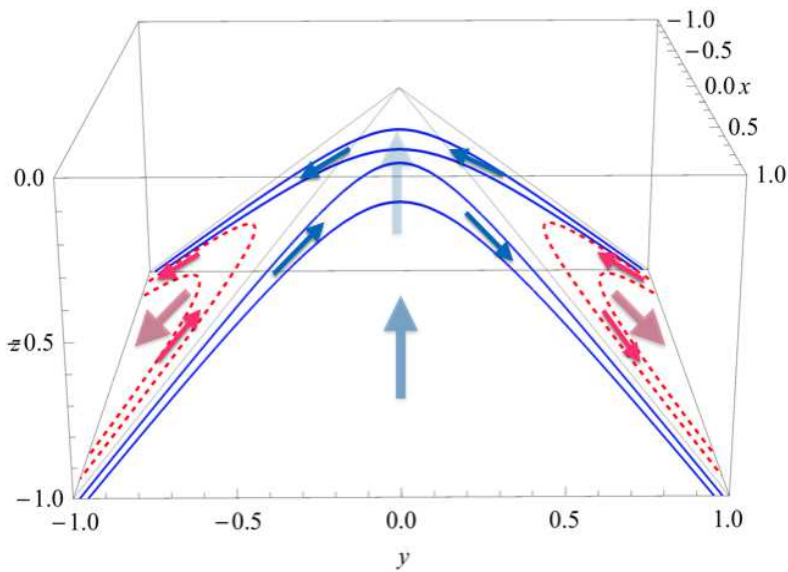}
\end{center}
\caption{Sketch indicating the nature of the pyramid-reconnection process: the sense of the surviving vortex filaments (solid blue, with associated flux $\Gamma\!_{s}(\tau)$ and direction as indicated by the arrows) is such that they propagate upwards under their mutual interaction towards the apex of the prism, where they reconnect; the  reconnected vortex filaments  (dashed red, with associated flux $\Gamma\!_{r}(\tau)$) propagate downwards on the complementary faces of the prism away from the apex.}
\label{Fig_pyramid_reconnection}
\end{figure}

 The `pyramid-reconnection' process, as anticipated in MK19,  is represented schematically by the sketch of figure \ref{Fig_pyramid_reconnection}. As $\delta/s$ increases to 
O($1$) it is necessary to take account of the process of viscous vortex reconnection on the symmetry plane $x=0$.  This leads to a stripping away of some of the $y$-component of vorticity at the tipping points in each half-space $x<0$ and $x>0$.  Thus, the initial circulation $-\Gamma$ in the vortex ${\cal V}_{1}$ centred on the curve
$C_{1}$ in the half-space $x<0$ branches into two circulations: `surviving circulation'  $-\Gamma\!_{s}(\tau)$ (blue, solid) and `reconnected circulation'
$-\Gamma\!_{r}(\tau)$ (red, dashed), with $\Gamma\!_s(\tau)+\Gamma\!_{r}(\tau)=\Gamma$ (and similarly of course for the vortex ${\cal V}_{2}$ in $x>0$). The arrows in Figure \ref{Fig_pyramid_reconnection} indicate the compatible directions of vorticity $\bfomega$ in the surviving and reconnected domains.  The reconnected circulation is oriented so that it falls away from the interaction region; this presumably exerts a mild braking effect on the continuing progress of the surviving circulation towards the incipient singularity. Here, in exploratory vein, we shall neglect this braking effect, which could however be included in an improved model. 

These branching and falling-away processes are essentially the same as  the `bridging' and `stripping' processes envisaged  by Melander \& Hussain (1989)  and Hussain \& Duraisami (2011). The processes have been recently described in similar terms by Kerr (2018), through DNS computations on vortex reconnection for a trefoil vortex and for the more conventional situation of perturbed anti-parallel vortices. For the latter case, Kerr describes an ``exchange of circulations", similar to the viscous destruction of surviving circulation and its replacement by reconnected circulation, a process that we now analyse. We note that Hussain \& Duraisami (2011) describe vortex reconnection at vortex Reynolds numbers up to 7000; the principal vortices remain coherent both before and after reconnection, and they observe that ``viscous reconnection is never complete, leaving behind a part of the initial tubes as threads, which then undergo successive reconnections". Flattening of the primary viscous core cross-sections is not apparent in this study, but flattening does occur in the interaction of the bridging threads after reconnection.  We cannot capture these `detritus effects' in our analytical model, which is primarily concerned with the build-up to reconnection and to the time when the stretched vorticity is maximal.

The rate of decrease of the surviving circulation $\Gamma\!_s(\tau)$ 
results from viscous diffusion across the plane $x=0$, and is given in dimensionless form by
\be\label{Gamma_s_0}
\frac{\tn{d}\Gamma\!_{s}}{\tn{d}\tau}=\frac{\tn{d}}{\tn{d}\tau}\oint_{C_{-}}{\bf{v}}\,\cdot\tn{d}{\bf{x}}=\epsilon\oint_{C_{-}}\nabla^{2}{\bf{v}}\,\cdot\tn{d}{\bf{x}}\,,
\ee
where $C_{-}$ is the closed circuit consisting of the $z$-axis ($x=y=0$) and the semi-circle at infinity in the half-plane $y=0, \,x<0$.  (Note that when $\epsilon=0$ this equation is an expression of Kelvin's circulation theorem for the stationary circuit $C_{-}$; when $\epsilon>0$, we need merely include the term $\epsilon \nabla^{2} \bf{v}$, from the Navier-Stokes equation.) It is evidently only the contribution from  the $z$-axis here that is non-zero, and this gives
\be\label{Gamma_s}
\frac{\tn{d}\Gamma\!_{s}}{\tn{d}\tau}=
\epsilon\int_{-\infty}^\infty \left.\left(\frac{\partial^2 \varv_{z}}{\partial x^2}+\frac{\partial^2 \varv_{z}}{\partial z^2}\right)\right|_{x=0}\tn{d}z=
\epsilon\int_{-\infty}^\infty \left.\frac{\partial^2 \varv_{z}}{\partial x^2}\right|_{x=0}\tn{d}z,
\ee
(the second term integrating to zero because $\partial \varv_{z}/\partial z=0$ at $z=\pm\infty$.)
 \begin{figure}
\begin{center}
\begin{minipage}{0.99\textwidth}
\hspace*{12pt}
\includegraphics[width=0.45\textwidth, trim=0mm 0mm 0mm 0mm]{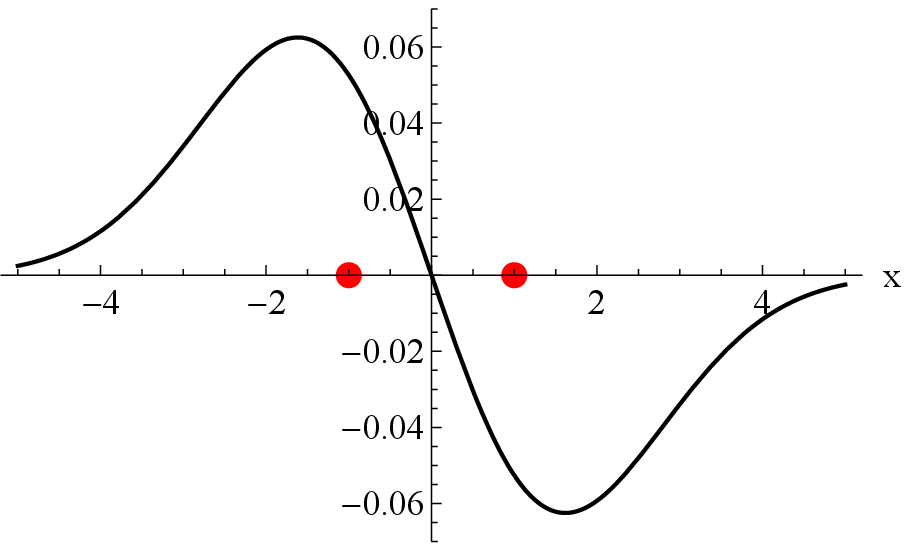}
\hspace*{30pt}
\includegraphics[width=0.45\textwidth,  trim=0mm 0mm 0mm 0mm]{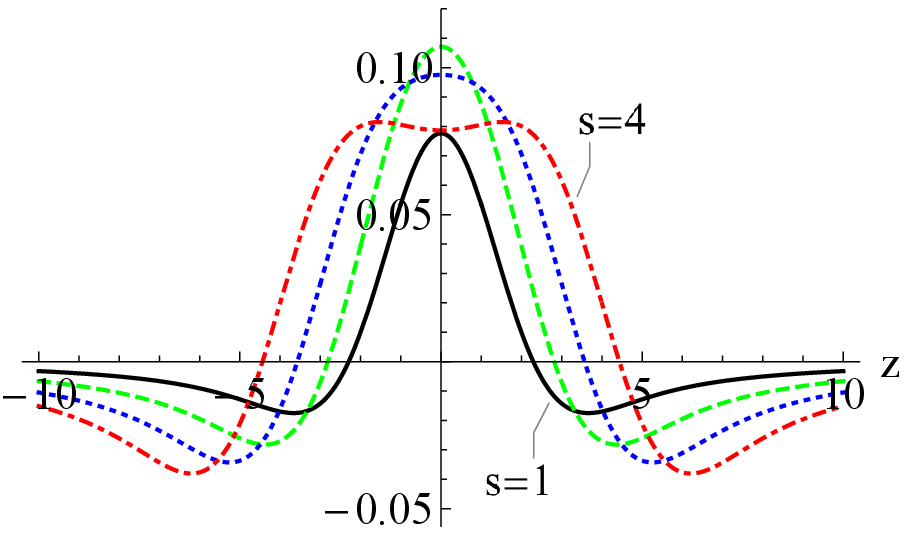}
\end{minipage}
\end{center}
\vskip 2mm
\hskip 30mm (a) \hskip 65mm (b) 
\vskip 2mm
\caption{(a) The vorticity $\omega_{y}(x,\,0,\tau)/\Gamma\!_{s}(\tau)$ given by (\ref{omega_two_vortices}), at a time $\tau$ at which  $s=1, \delta/s=0.9456$ (the limiting value stated above); the position of the vortex centres is indicated by the two red dots; (b) the  velocity $\varv_{z}(x,\,0,\tau])/\Gamma\!_{s}(\tau)$ given by (\ref{vel_z}) for $s=1$ (solid), 2 (green, dashed, 3 (blue, dotted) and 4 (red, dot-dashed).}
\label{Fig_04_omega_y_v}
\end{figure}

Now neglecting the core deformation, the vorticity field on the plane $y=0$ at time $\tau$ is that of two vortices of Gaussian structure, with the single component 
\be\label{omega_two_vortices}
\omega_{y}(x,\,z,\,\tau)=\frac{\Gamma\!_{s}(\tau)}{4\pi\delta^2}
\left(\exp{\left[-\frac{(x+s)^2+z^2}{4\delta^2}\right]}-
\exp{\left[-\frac{(x-s)^2+z^2}{4\delta^2}\right]}\,\right)\,,
\ee
as sketched in figure \ref{Fig_04_omega_y_v}a for $s=1, \delta=0.9456$. The corresponding $z$-component of velocity is
\be\label{vel_z}
\varv_{z}(x,z,\tau)\!=\!\frac{\Gamma\!_{s}(\tau)} {2\pi}\!\left\{\!
\frac{x+s}{{(x\!+\!s)^2\!+\!z^2}}\left(\!1\!-\!\exp{\left[-\frac{(x\!+\!s)^2\!+\!z^2}{4\delta^2}\right]}\right)-\frac{x-s}{{(x\!-\!s)^2\!+\!z^2}}\left(\!1\!-\!\exp{\left[-\frac{(x\!-\!s)^2\!+\!z^2}{4\delta^2}\right]}\right)\right\},
\ee
(figure \ref{Fig_04_omega_y_v}b, for  $s=1, 2, 3, 4$ and fixed $\delta=0.9456$), 
from which we may calculate 
\bea\label{v_zz}
\left.\frac{\partial^2 \varv_{z}}{\partial x^2}\right|_{x=0}\!\!=\!&-&\frac{s^{5}\,\Gamma\!_{s}(\tau)\left[\left(s^6 \!+\! 2s^{4}z^{2}\!-\!6z^{4}\delta^2 \!-\!24z^{2}\delta^2  \!+\! s^2 z^4 \!-\!4s^2 z^2 \delta^2  \!+\! 8s^2\delta^4\right)\exp{\left[-(s^2 \!+\! z^2)/4\delta^2\right]}\right]}{2\pi(s^2+z^2)^{3}\delta^2}\nonumber\\  &-& \frac{ 2 s\,\Gamma\!_{s}(\tau)(3z^2-s^2)}{\pi(s^2+z^2)^{3}}\,.
\eea
Here, the second term is just what survives in the `point-vortex' limit $\delta\rightarrow 0$, and it integrates to zero over  the range $z \in (-\infty$, $\infty)$.
On this integration, the first term contributes the result
\be\label{int_dd_vel}
\int_{-\infty}^\infty \left.\frac{\partial^2 \varv_{z}}{\partial x^2}\right|_{x=0}\tn{d}z=
-\frac{s\,\Gamma\!_{s}(\tau)}{2\surd{\pi}\,\delta^3}\exp[-s^2/4\delta^2]\,,
\ee
and it  follows from (\ref{Gamma_s})  that
\be\label{Gamma_s2}
\frac{\tn{d}\Gamma\!_{s}}{\tn{d}\tau}=-\epsilon\,\frac{s\,\Gamma\!_{s}(\tau)}{2\surd{\pi}\,\delta^3}\exp[-s^2/4\delta^2]\,.
\ee

The above neglect of core deformation calls for comment. There is evidence from DNS,  such as in the recent investigations of  McKeown et al. (2018)  and Kerr (2018) and in earlier studies, that there is a flattening of the vortex cores during the interaction process when $\delta/s$ increases to O$(1)$. We have argued in MK19 that this flattening should decrease with increasing Reynolds number, on the grounds that the vortices are then spinning so rapidly that they experience an effectively axisymmetric strain near the tipping points. Nevertheless, the curvature of the vortices at the tipping points may affect this conclusion, a possibility that calls for further analysis.  A modest degree of core flattening could easily be taken into account through simple modification of the vorticity structure (\ref{omega_two_vortices}) assumed above; this would change the details of the above calculation, but is unlikely to modify the result (\ref{Gamma_s2}) in any fundamental way.  We therefore adopt this result as it stands. It is important to note that this result is firmly based on (\ref{Gamma_s_0}), which itself is a direct consequence of the Navier-Stokes equation.

 During the very short reconnection phase, we thus need to replace $\Gamma$ by 
$\Gamma\!_{s}(\tau)$ in calculating the instantaneous velocity, rate of stretching, and rate-of-change of curvature at the tipping points.  This simply means that, defining 
$\gamma(\tau)\equiv \Gamma\!_{s}(\tau)/\Gamma$, we have to replace (\ref{system1}) by
\be\label{system3}
\frac{\tn{d} s}{\tn{d}{ \tau}}=-\frac{\gamma\,\kappa\cos\alpha}{4\pi}\left[\log\left(\frac{s}{\delta}\right)+\beta_{1}\right] \,,\quad \frac{\tn{d} \kappa}{\tn{d}{\tau}}=\frac{\gamma\,\kappa\cos\alpha\sin\alpha}{4\pi {s}^2}\,,\quad \frac{\tn{d} \,\delta^2}{\tn{d}{\tau}}=\epsilon -\frac{\gamma\,\kappa\cos\alpha}{4\pi {s}} \,\delta^{2}\,,
\ee
which, from (\ref{Gamma_s2}), must now be coupled with
\be\label{gamma}
\frac{\tn{d}\gamma}{\tn{d}\tau}=-\epsilon\,\frac{s\,\gamma}{2\surd{\pi}\,\delta^3}\exp[-s^2/4\delta^2]\,.
\ee
 We may now integrate this 4th-order dynamical system with the initial conditions (\ref{initial}), together with $\gamma(0)=1$.

\section{Integration of the system (\ref{system3}, \ref{gamma}) }
We have integrated this system, using Mathematica with 56-point precision,  
with a view to determining the variation of 
$\omega(\tau)\equiv\gamma(\tau)/\delta(\tau)^2$; \,this is a measure of the maximum vorticity at the centre of the surviving vortex core.  The function $\omega(\tau)/\omega(0)$ then represents the amplification of vorticity.  There are two competing effects here:  $\delta^2$ decreases, at least initially, towards zero, but $\gamma$ also decreases due to the reconnection process.

\begin{figure}
\begin{center}
\begin{minipage}{0.99\textwidth}
\hspace*{2pt}
\includegraphics[width=0.48\textwidth, trim=0mm 0mm 0mm 0mm]{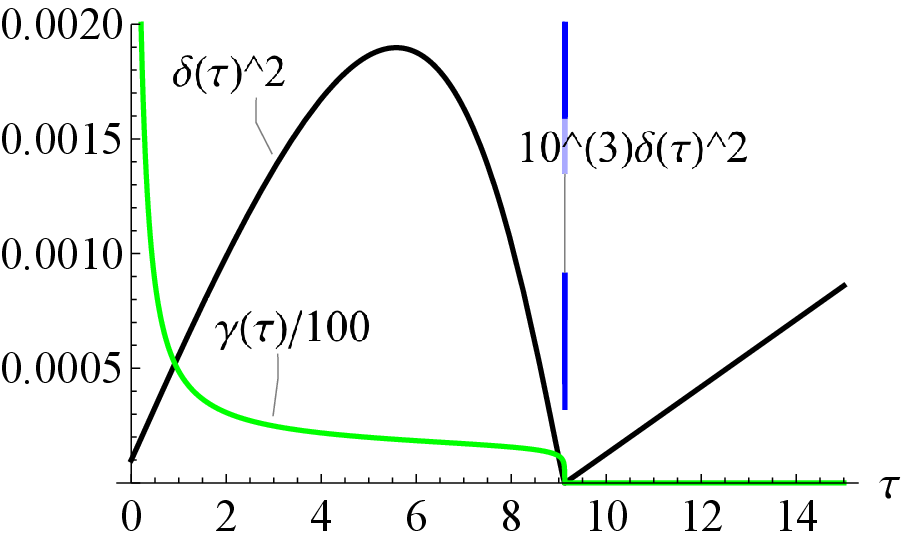}
\hspace*{3pt}
\includegraphics[width=0.48\textwidth,  trim=0mm 0mm 0mm 0mm]{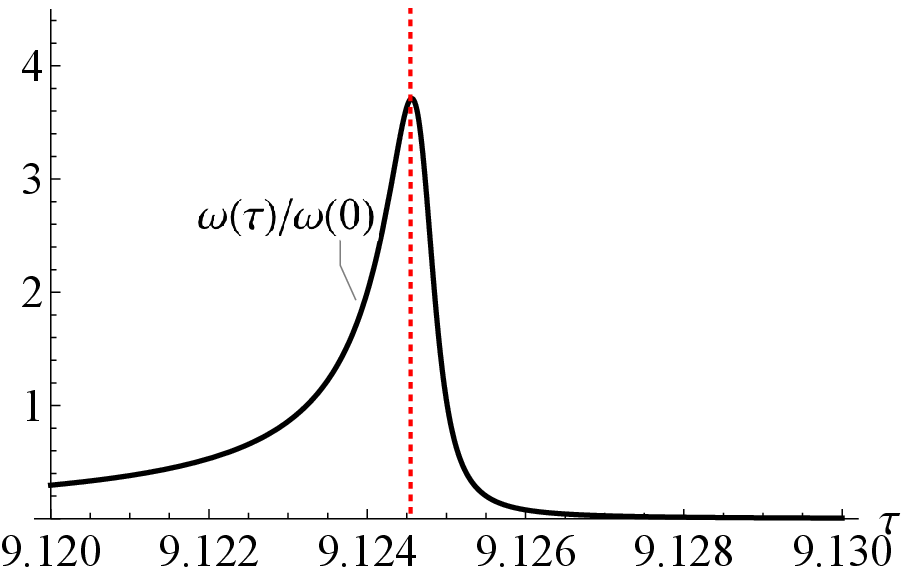}
\end{minipage}
\end{center}
\vskip 2mm
\hskip 30mm (a) \hskip 65mm (b) 
\vskip 2mm
\caption{Evolution governed by (\ref{system3}) with $\epsilon=1/2000$: (a) $\delta(\tau)^2$ (black) which appears to fall to zero at $\tau=\tau_c\approx 9.12455$, but does not quite reach zero as shown by the expanded curve of $10^3\delta(\tau)^2$ (blue); $\gamma(\tau)/100$ (green) decreases to about 0.0002 at $\tau\approx \tau_c$ then falls rapidly to very near zero; (b) the corresponding variation of vorticity amplification $\omega(\tau)/\omega(0)$ in the neigbourhood of $\tau_c$, where it reaches a maximum of approximately 3.7. }
\label{Fig_05_2000}
\end{figure}

Figure \ref{Fig_05_2000}a shows results for $\epsilon=1/2000$, i.e. $R_\Gamma =2000$. This shows that $\delta^2$ increases initially due to viscous diffusion, but then vortex stretching dominates, causing a decrease to very near (but not quite) zero; in fact, $\delta^2$ reaches a minimum of $2.67412\times10^{-8}$ at $\tau=\tau_c\approx 9.12455$, and then for $\tau\gtrsim  \tau_c$, increases almost linearly due to viscous diffusion.  The variation of $\gamma(\tau)/100$ is shown, the factor $1/100$ being introduced to make the variation more visible; $\gamma$ decreases from 1 to about 0.02 as $\tau$ approaches $\tau_c$ and then drops very rapidly to zero, indicating rapid completion of the reconnection process.  
Figure \ref{Fig_05_2000}b shows the corresponding variation of $\omega(\tau)/\omega(0)$, which first increases due to the decrease of $\delta^2$, but reaches a modest maximum of $~3.711$. Thus, there is certainly no singularity for this value of $R_\Gamma$.

\begin{figure}
\begin{center}
\begin{minipage}{0.99\textwidth}
\hspace*{2pt}
\includegraphics[width=0.48\textwidth, trim=0mm 0mm 0mm 0mm]{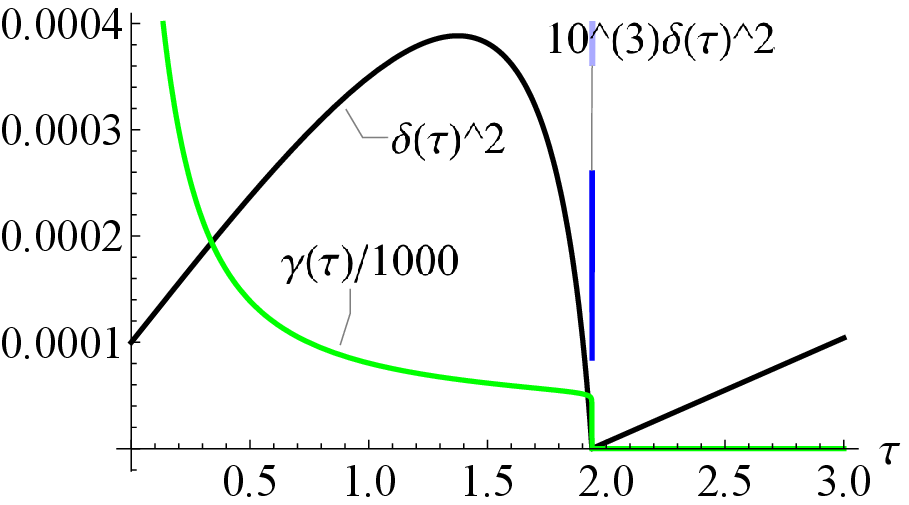}
\hspace*{3pt}
\includegraphics[width=0.48\textwidth,  trim=0mm 0mm 0mm 0mm]{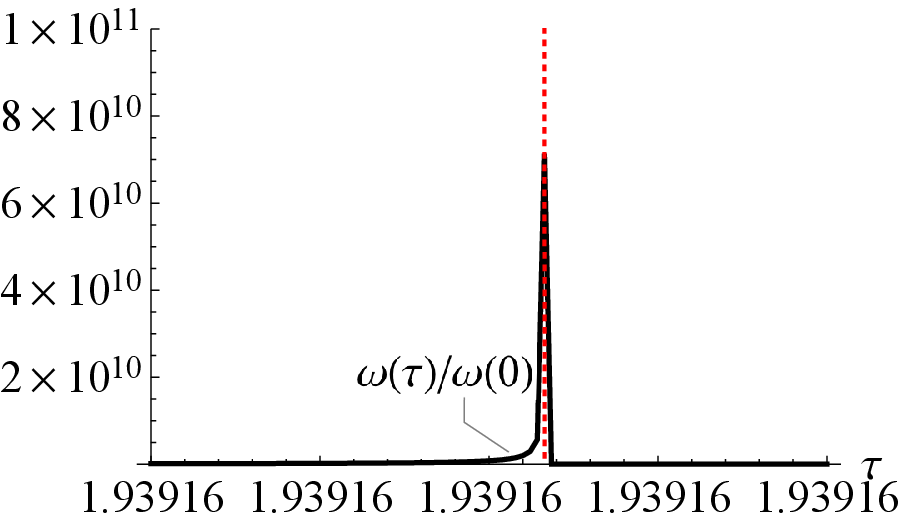}
\end{minipage}
\end{center}
\vskip 2mm
\hskip 30mm (a) \hskip 65mm (b) 
\vskip 2mm
\caption{Same as figure \ref{Fig_05_2000}, but with $\epsilon=1/3000$.}
\label{Fig_06_3000}
\end{figure}

Figure \ref{Fig_06_3000} shows corresponding results with $\epsilon=1/3000$.  Again, $\delta^2$ increases initially but then decreases to near zero at $\tau=\tau_c\approx 1.9391619710016794917768$, but still just fails to reach zero; 
in fact, $\delta^2$ evaluates to $2.7278\times10^{-24}$ at $\tau=\tau_c$!  \,$\gamma(\tau)$ has decreased to $\sim 0.05$ at $\tau=1.93$ and drops to $0.000451$ at $\tau=\tau_c$. Now, $\omega(\tau)/\omega(0)$ (figure \ref{Fig_06_3000}b) shows a very sharp peak just before $\tau=\tau_c$ where, in fact $\omega(\tau_c)/\omega(0)$ evaluates to $2.425\times10^{16}$, a huge increase caused by vortex stretching, but still not a singularity.

\begin{figure}
\begin{center}
\begin{minipage}{0.99\textwidth}
\hspace*{2pt}
(a)\includegraphics[width=0.47\textwidth, trim=0mm 0mm 0mm 0mm]{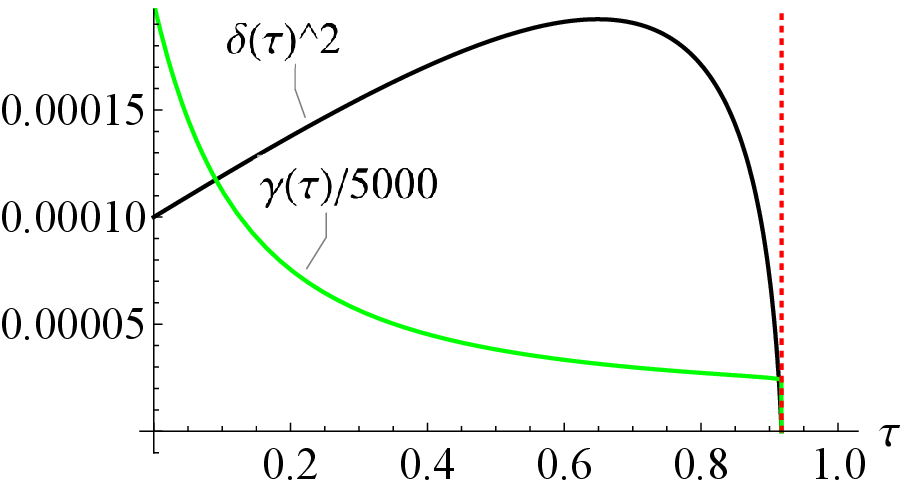}
\hspace*{10pt}
(b)\includegraphics[width=0.47\textwidth,  trim=0mm 0mm 0mm 0mm]{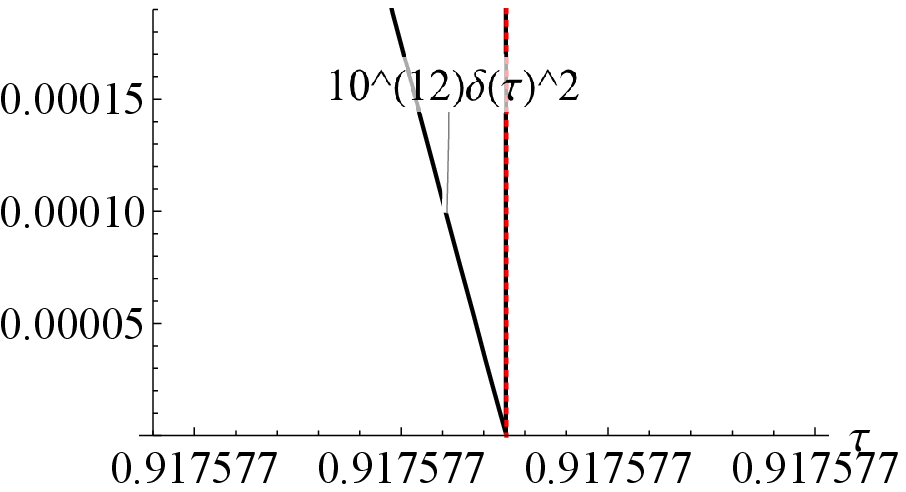}
\\
\vskip 3mm
\hspace*{2pt}
(c)\includegraphics[width=0.47\textwidth, trim=0mm 0mm 0mm 0mm]{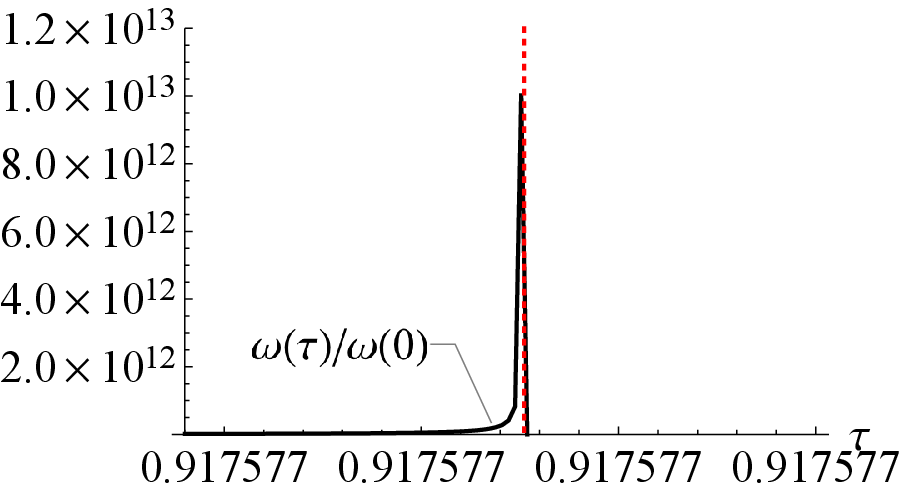}
\hspace*{10pt}
(d)\includegraphics[width=0.47\textwidth,  trim=0mm 0mm 0mm 0mm]{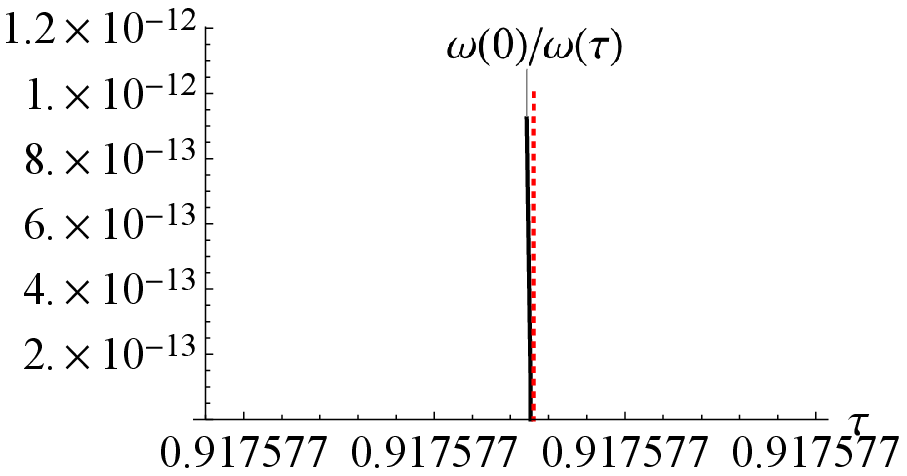}
\\
\vskip 3mm
\hspace*{2pt}
(e)\includegraphics[width=0.48\textwidth, trim=0mm 0mm 0mm 0mm]{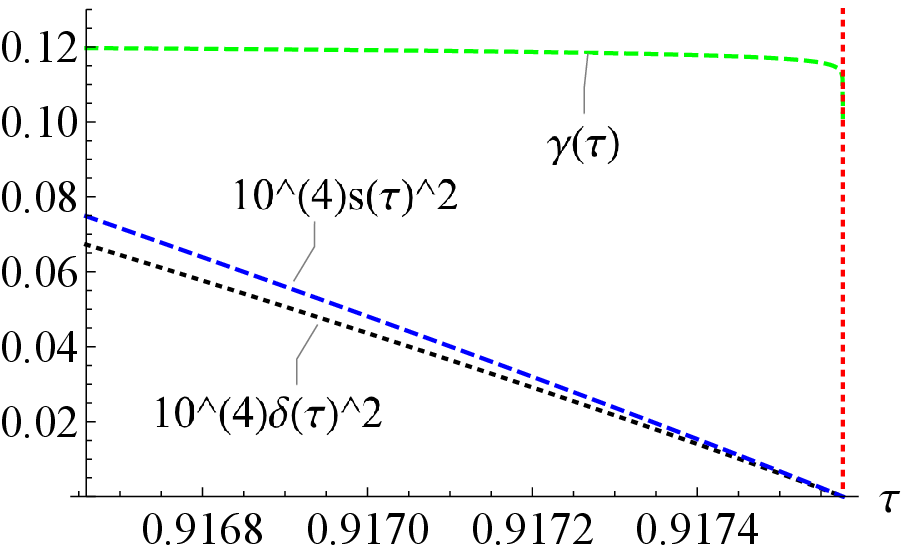}
\hspace*{3pt}
(f)\,\,\includegraphics[width=0.48\textwidth,  trim=0mm 0mm 0mm 0mm]{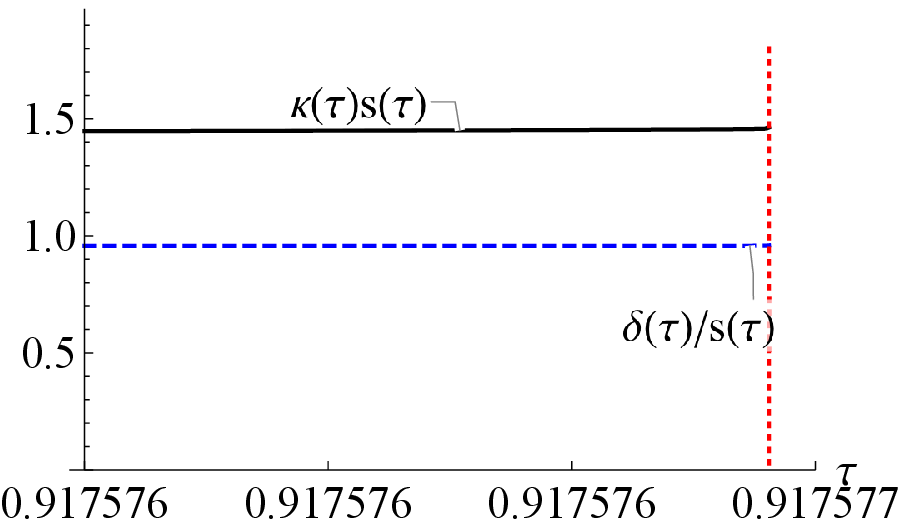}
\end{minipage}
\end{center}
\vskip 2mm
\vskip 2mm
\caption{ Evolution for $\epsilon=1/4000$, and for $0<\tau<\tau_c\approx0.917\,576\,564\,746\,480\dots$ (marked by red dotted line; (a) $\delta^{2}(\tau)$ and $\gamma(\tau)$ both appear to fall to zero at $\tau=\tau_c$; zoom of $10^{12}\delta^{2}(\tau)$ very near $\tau_c$; (c) and (d) $\omega(\tau)/\omega(0)$ and $\omega(0)/\omega(\tau)$ near $\tau_c$, as well as can be resolved; (e) and (f) other relevant variables near $\tau_c$, as well as can be resolved.}
\label{Fig_06_4000}
\end{figure}

Consider now the situation when $\epsilon=1/4000$ (i.e. $R_\Gamma=4000$).  Here, Mathematica  stalls at $\tau=\tau_c \approx 0.917 \,576\, 564 \,746\, 48$. The reason is that now $\delta^2$  falls so close to zero that equation (\ref{gamma}) runs into an apparent singularity.  Figure \ref{Fig_06_4000}a shows $\delta^2(\tau)$ and $\gamma(\tau)/5000$ (again the factor $1/5000$ is included to make the comparison clear), showing that, while $\delta^2$ seems to go to zero at the apparent singularity time $\tau=\tau_c$ (marked by the dotted line), $\gamma$ is still finite;  figure \ref{Fig_06_4000}b shows $\delta^2$ expanded by a factor $10^{12}$ in the immediate neighbourhood of $\tau=\tau_c$ ($ 0.917 \,576\, 564\, 746\,39 <\tau$ $<0.917\, 576\, 564 \,746\, 55$); $\delta(\tau_c)^2$ is certainly extremely small if not exactly zero.
Figure \ref{Fig_06_4000}c  shows (over the same exceedingly small range of $\tau$) that $\omega(\tau)/\omega(0)$ increases to a very large value near $\tau_c$; but is this really infinite?  Figure \ref{Fig_06_4000}d shows the inverse function $\omega(0)/\omega(\tau)$, which does indeed appear to be zero at this point as far as this computation can determine. 
\begin{figure}
\begin{center}
\begin{minipage}{0.99\textwidth}
\hspace*{0pt}
(a)\,\,\includegraphics[width=0.43\textwidth, trim=0mm 0mm 0mm 0mm]{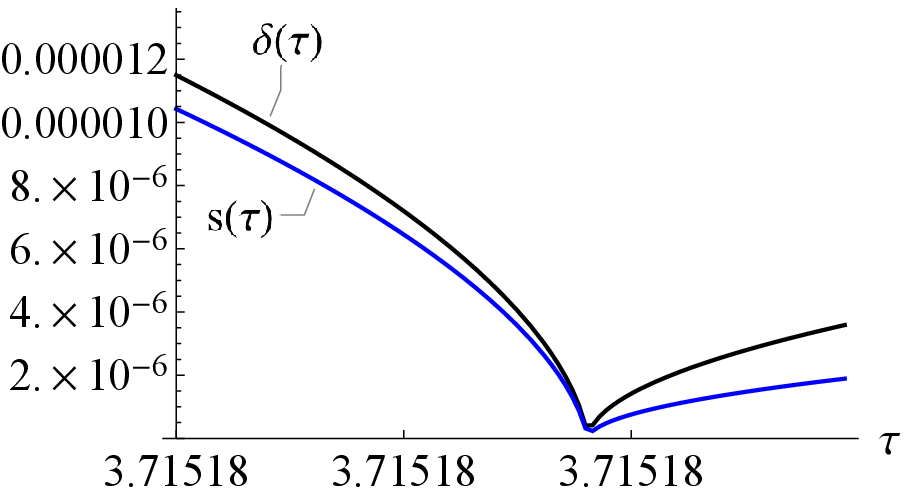}
\hspace*{10pt}
(b)\,\,\includegraphics[width=0.43\textwidth, trim=0mm 0mm 0mm 0mm]{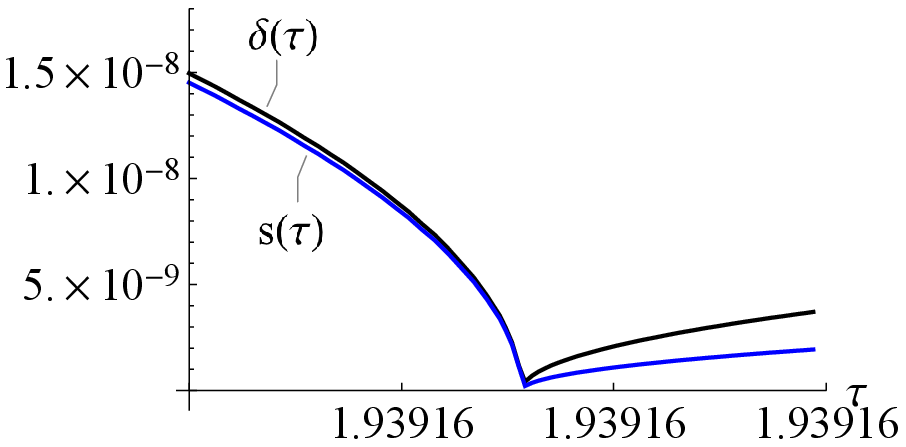}
\hspace*{10pt}\\
(c)\,\,\includegraphics[width=0.43\textwidth, trim=0mm 0mm 0mm 0mm]{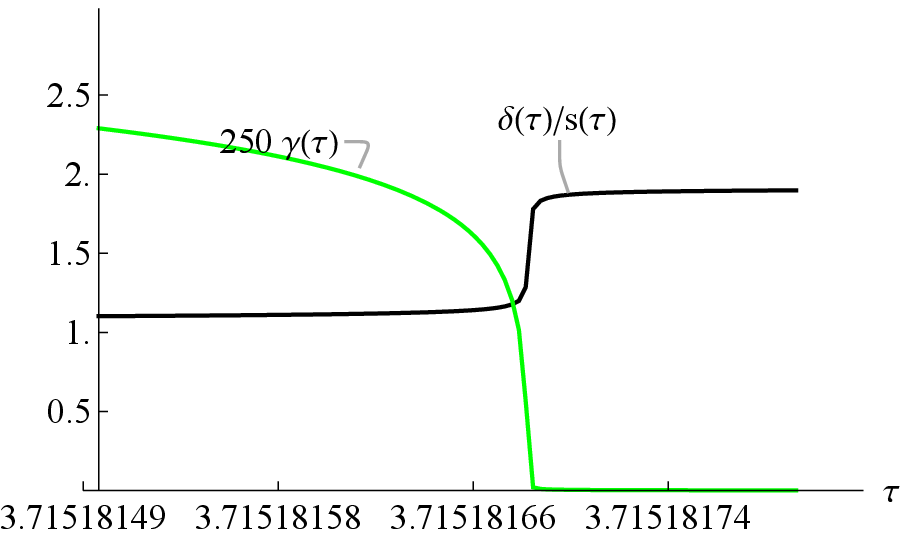}
\hspace*{10pt}
(d)\,\,\includegraphics[width=0.43\textwidth,  trim=0mm 0mm 0mm 0mm]{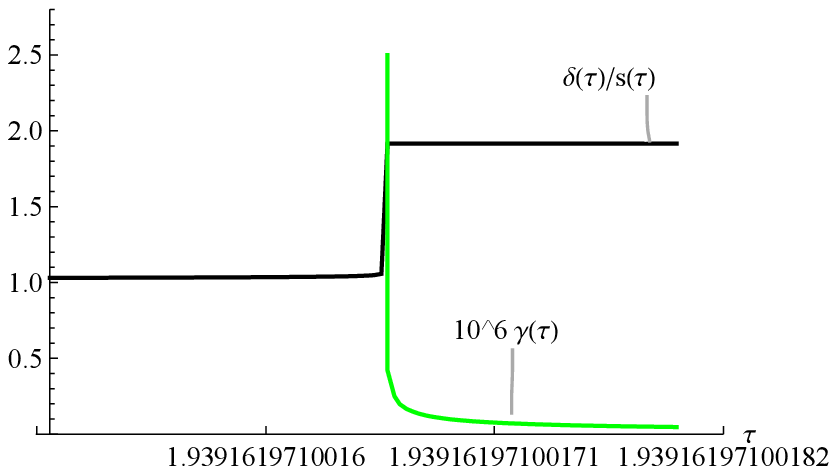}
\end{minipage}
\end{center}
\vskip 2mm
\caption{Behaviour of  $\delta (\tau),\, s(\tau)$ and the ratio $\delta (\tau) /s(\tau)$ during the very small time-interval when $\gamma (\tau)$ (shown in green) decreases rapidly to near zero;  (a) and (c) $\epsilon=1/2500$; the abscissa runs from 3.71518149 to 3.7151818 and $\gamma (3.72) = 3.21945\times 10^{-8}$; (b) and (d)  $\epsilon=1/3000$; the abscissa runs from 1.93916197100149 to 1.93916197100182 and $\gamma (1.194) = 6.05673\times 10^{-13}$. In both cases, $\gamma(\tau)$ falls to an extremely low value, but not exactly zero; the remnant weak vorticity expands solely under the action of viscous diffusion.}
\label{Fig_gds}
\end{figure}

\section{Asymptotic behaviour for small $\epsilon$}
A necessary condition for a finite-time singularity of vorticity $\omega(\tau)\equiv\gamma(\tau)/\delta(\tau)^2$ is obviously that $\delta(\tau)^2$ should decrease to exactly zero at some finite time. In order to further investigate this possibility, we consider the $\epsilon$-dependence of $\delta(\tau)^2_{min}$ as $\epsilon$ decreases towards zero from $1/2000$.  For each $\epsilon$, we first locate the time $\tau_c$  at which $\tn{d} \,\delta^2/\tn{d}{\tau}\,\,(=\epsilon -(\gamma\,\kappa\cos\alpha/4\pi {s}) \,\delta^{2})$  is zero, and then evaluate the corresponding minimum value of $\delta(\tau)^2$; this requires a very sharp refinement process to  accurately identify the time $\tau_c$ at which this minimum actually occurs.

 \begin{table}\label{table_delta}
   \begin{center} 
   \begin{tabular}{ clcc}\hline 
$\epsilon$ & $\,\,\,\,\,\,\,\,\,\,\,\,\,\tau_c$ & $\delta^2_{min}$& \,\,\,\,\,$\gamma(\tau_c)$  \\ \hline
1/2000 &9.12478705   &$2.67412 \times10^{-8}$ &\,\,\,\,\,0.000677  \\ 
1/2250 &5.64196432008   &$8.34641 \times10^{-11}$&\,\,\,\,\,0.000602  \\
1/2500 &3.71518168478948 &$ 3.69041 \times10^{-14}$&\,\,\,\,\,0.000542  \\ 
1/2750 &2.60587589889538 & $1.37078 \times10^{-18}$&\,\,\,\,\,0.000501 \\ 
1/3000 & 1.9391619710016794917768 & $2.72780\times 10^{-24}$&\,\,\,\,\,0.000451  \\
1/3050 & 1.83975957787485971515631271&$1.34992\times 10^{-25}$&\,\,\,\,\,0.000444 \\
1/3100 &1.748954708554952698937886817 &$5.85955\times 10^{-27}$&\,\,\,\,\,0.000437 \\
1/3200 &1.5896543686535978867264670724 &$7.41570\times 10^{-30}$&\,\,\,\,\,0.000423\\
1/3300 &1.4552611229433966364089007674807  &$5.49450\times 10^{-33}$&\,\,\,\,\,0.000410\\ 
1/3400 &1.341068010282669618710465713447975557 &$2.37870\times 10^{-36}$&\,\,\,\,\,0.000399\\
  \hline\hline
\end{tabular}
\end{center}
 \caption{Dependence of the minimum attained by $\delta^{2}(\tau)$; as $\epsilon$ decreases the time $\tau_c$ at which this minimum is attained has to be determined with ever increasing accuracy as indicated by the number of decimal points.}
 \label{table_delta}
\end{table}

\begin{figure}
\begin{center}
\begin{minipage}{0.99\textwidth}
\hspace*{0pt}
(a)\includegraphics[width=0.45\textwidth, trim=0mm 0mm 0mm 0mm]{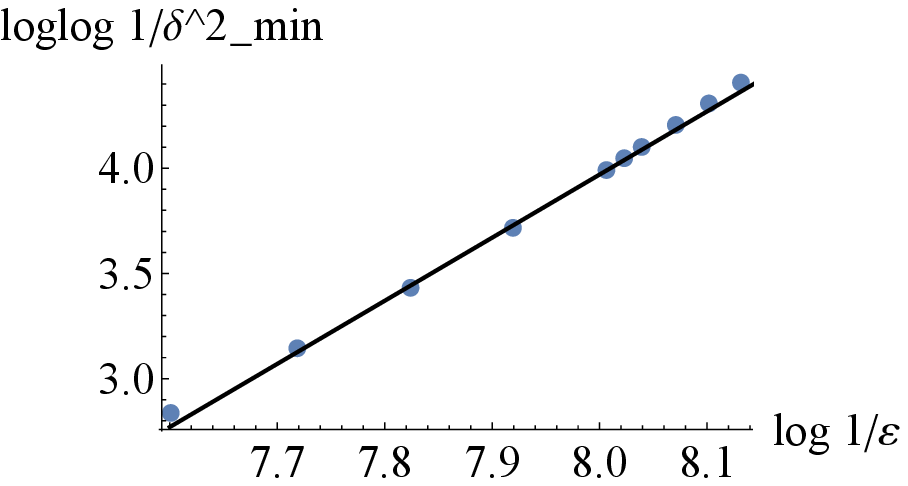}
\hspace*{10pt}
(b)\includegraphics[width=0.45\textwidth,  trim=0mm 0mm 0mm 0mm]{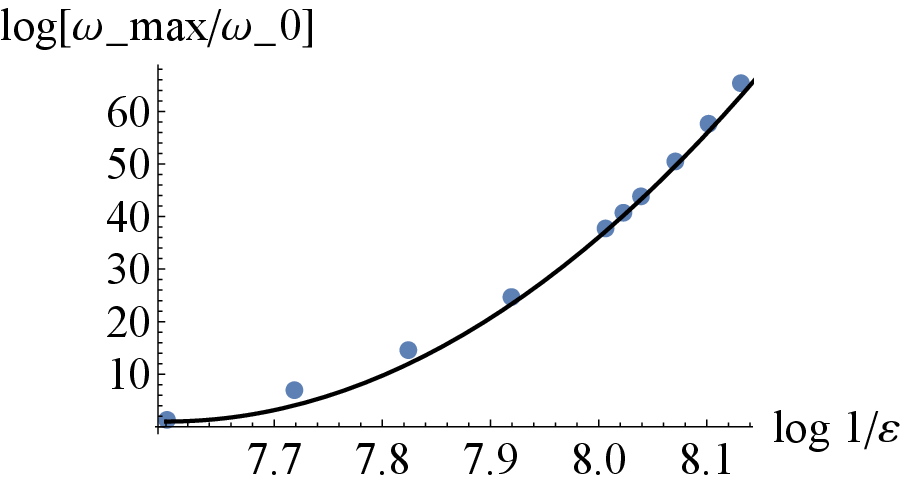}\\
\vskip 2mm
\hspace*{0pt}
(c)\includegraphics[width=0.45\textwidth,  trim=0mm 0mm 0mm 0mm]{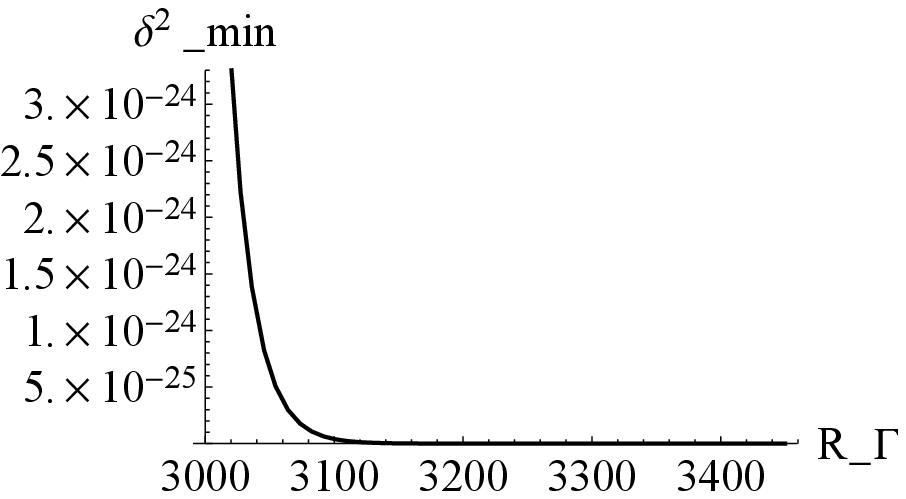}
\hspace*{10pt}
(d)\includegraphics[width=0.45\textwidth,  trim=0mm 0mm 0mm 0mm]{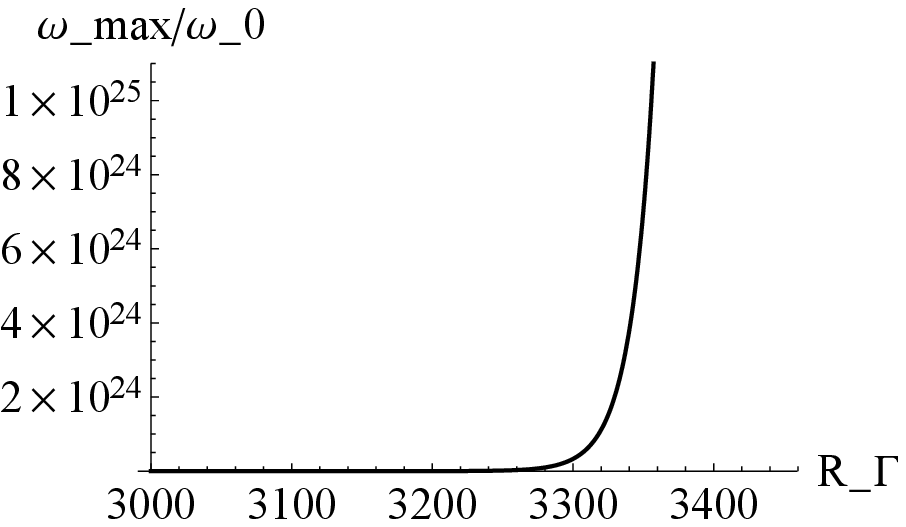}
\end{minipage}
\end{center}
\vskip 2mm
\caption{(a) Plot of the data of table \ref{table_delta}, fitted by the straight line (\ref{straight_line}); (b) same for the data of table \ref{table_omega},fitted by the quadratic curve (\ref{quadratic}); (c) and (d) inferred dependence of $\delta^{2}_{min}$ and $\omega_{max}/\omega_{0}$ on vortex Reynolds number $R_{\Gamma}$.}
\label{Fig_loglogs}
\end{figure}

 \begin{table}\label{table_omega}
   \begin{center} 
   \begin{tabular}{ cllc}\hline 
$\epsilon$ & $\,\,\,\,\,\,\,\,\,\,\,\,\,\tau_m$ & $\omega_{max}/\omega_{0}$ & \,\,\,\,\,$ \gamma(\tau_m)$ \\ \hline
1/2000 &9.12456266  &$3.711$   &\,\,\,\,\,0.00139\\
1/2250 &5.64196352000   &$1057$&\,\,\,\,\,0.00125  \\
1/2500 &3.71518168440194&$ 2.151 \times10^{6}$&\,\,\,\,\,0.00112 \\ 
1/2750 &2.605875898895364555 & $5.266 \times10^{10}$&\,\,\,\,\,0.00101 \\ 
1/3000 & 1.9391619710016794917425 & $2.425\times 10^{16}$&\,\,\,\,\,\,\,0.000930 \\
1/3050 &1.8397595778748597151545&$4.820\times 10^{17}$&\,\,\,\,\,\,\,0.000910\\
1/3100 &1.748954708554952698937812055 &$1.092\times 10^{19}$&\,\,\,\,\,\,\,0.000889 \\
1/3200 &1.589654368653597886726466973 &$8.363\times 10^{21}$&\,\,\,\,\,\,\,0.000871\\
1/3300 &1.455261122943396636408900767405 &$1.095\times 10^{25}$& \,\,\,\,\, 0.000843\\
1/3400 &1.341068010282669618710465713447942 &$2.454\times 10^{28}$& \,\,\,\,\,\,0.000800\\
  \hline\hline
\end{tabular}
\end{center}
 \caption{Dependence of the maximum attained by $\omega(\tau)/\omega(0)$;  the time $\tau_m$ at which this maximum is attained is slightly less than $\tau_c$, and $\gamma(\tau_m)$ is correspondingly slightly greater than $\gamma(\tau_c)$.}
 \label{table_omega}
\end{table}

The results are shown in table \ref{table_delta} and in the  corresponding plot of $\log{\log {\delta^{-2}|_{max}}}$ as a function of $\log{\epsilon^{-1}}$  (figure \ref{Fig_loglogs}a). The straight line fit provides the relationship
\be\label{straight_line}
\log{\log {\delta^{-2}|_{max}}}\sim 3 \log{\epsilon^{-1}}-20.03,
\ee
or equivalently
\be\label{asym_d}
{\delta^{2}|_{min}}\sim \exp{\left[-53 \left(R_\Gamma/3000\right)^{3}\right]}\,.
\ee
This does indeed indicate an extremely rapid decrease of $\delta^{2}|_{min}$  as 
$R_\Gamma=\epsilon^{-1}$ increases beyond about $3000$, as shown in figure \ref{Fig_loglogs}c ; for example, when $R_\Gamma=4000$, it gives $\delta^{2}|_{min}=2.75\times 10^{-55}$, which  appears to be zero  in figure \ref{Fig_06_4000}; but in fact the asymptotic relation (\ref{asym_d}), if valid as $\epsilon\rightarrow 0$, indicates that for any $\epsilon>0$, $\delta^{2}|_{min}$ does not vanish, and we do not therefore have a strict mathematical singularity. 

The behaviour of $\delta/s$ near $\tau_{c}$ is not well resolved in Figure  \ref{Fig_06_4000}(f), despite the extreme expansion of scale in this region. Figure \ref{Fig_gds} shows the situation for two larger values of $\epsilon\, (1/2500$ and $1/3000)$, which are more easily resolvable; these show that $\delta/s$ actually increases during the very short time-interval when $\gamma$ decreases rapidly to very near zero. This increase is however bounded: it rises to near 2 in both cases, and remains at this level in the subsequent purely viscous diffusion of the remnant vorticity; $\gamma$ is not exactly zero for $\tau>\tau_{c}$, as indicated in the figure caption.

Although we do not have a strict mathematical singularity, we may nevertheless seek to determine the maximum vorticity attained, as a function of $\epsilon$.  Computed values are shown in table \ref{table_omega}, and plotted in figure \ref {Fig_loglogs}(b) together with the closely fitting curve
\be\label{quadratic}
\log{[\omega_{max}/\omega_0]}\sim 1 + 220 ( \log{[1/\epsilon]} - 7.6)^2,
\ee
or equivalently
\be\label{quadratic_2}
\omega_{max}/\omega_0 \sim \exp{\left[1 + 220 \left(\log\left[R_{\Gamma}/2000\right]\right)^{2}\right]}\,.
\ee
Actually this formula underestimates the values computed for $R_\Gamma \gtrsim 3200$, and should  be regarded as a lower bound on $\omega_{max}/\omega_0$ for larger values of $R_\Gamma$.
The dependence (\ref{quadratic_2}), shown in figure \ref{Fig_loglogs}, shows a rapid increase for $R_\Gamma \gtrsim 3200$; and if we assume that this trend continues for larger $R_\Gamma$, then for $R_\Gamma=4000$ it gives $\omega_{max}/\omega_0 \sim 2\times 10^{46}$, and for  $R_\Gamma=5000$ it gives $\omega_{max}/\omega_0 \sim 4.5\times 10^{80}$!  We should note however that this peak of vorticity occurs when a large fraction of the original circulation has already reconnected (e.g. $\gamma(\tau_m)=0.008$ when $R_\Gamma=3400$ (table \ref{table_omega}); it is just the residual surviving vorticity that is apparently so intensely stretched.

\section{Conclusion}
We initially thought that the failure of Mathematica to resolve the behaviour of the system (\ref{system3}, \ref{gamma}) for $\epsilon=1/4000$ and smaller was an indication of a singularity at the time $\tau_c$ at which the programme stalled. However, computation of the dependence of 
$\delta^{2}_{min}$  on decreasing $\epsilon$ has revealed a smooth, although extremely sharp, cutoff in the range of 
$\epsilon$ in the interval $(1/3000,1/4000)$, such that, inevitably, there is a point beyond which Mathematica interprets $\delta^{2}_{min}$ as zero. Our inference however is that $\delta^{2}_{min}$ is nonzero for any $\epsilon>0$, although so small for $\epsilon\lesssim 1/4000$ that the core cross-sectional scale $\delta$ is  less than the inter-molecular scale below which the continuum hypothesis (on which the Navier-Stokes equation is based) is no longer justifiable. Moreover, the maximum vorticity amplification, attained just before this minimum of $\delta^{2}$ is reached, is so large that we may reasonably describe this as a physical, if not a mathematical, singularity.  The situation is reminiscent of the free-surface cusp singularity described by Jeong \& Moffatt (1992), again physical rather that mathematical, in which the radius of curvature at the `cusp' is O$(10^{-42})$ times the input scale of the problem.  The input scale of the present problem is the radius $R$ of the two vortices at time $\tau=0$,  so that if for example $R=10^{-1}$m, then even at $R_\Gamma=2750$, the figures of table \ref{table_delta} give $\delta\sim 10^{-10}$m, and we are down at the scale of the hydrogen atom. This is therefore quite evidently a physical singularity. In the light of this behaviour, the question ``can there be a finite-time singularity of the Navier-Stokes equations?" is  to some extent academic; but it continues to provide a legitimate academic challenge nonetheless!

Two physical processes have been ignored in the foregoing analysis: (i) deformation of the vortex cores during the reconnection process; and (ii) the possible braking effect exerted by the strands of reconnected vorticity on the continuing reconnection process.   These processes may limit the growth of vorticity (Hussain \& Duraisami 2011);  the fact that a singularity is averted even when they are ignored provides compelling evidence that, for the pyramid geometry considered here, a strict finite-time mathematical singularity does not occur, no matter how large the vortex Reynolds number may be.
\vskip 2mm
\noindent
YK acknowledges support from JSPS KAKENHI Grant Numbers JP18H04443,\,JP24247014,\,  JP16H01175. We acknowledge the  comments of three referees, which have led to improvements in the presentation.
\bibliographystyle{jfm}

\end{document}